\documentstyle[12pt]{article}

\global\arraycolsep=2pt
\begin{document}

\begin{titlepage}

\begin{flushright}
SLAC--PUB--6344\\
August 1993\\
T/E
\end{flushright}

\vspace{0.3cm}

\begin{center}
\Large\bf Short-Distance Expansion of Heavy-Light\\
Currents at Order 1/m$_{\bf Q}$
\end{center}

\vspace{0.4cm}

\begin{center}
Matthias Neubert\footnote{Supported by the Department of Energy
under contract DE-AC03-76SF00515.}\\
Stanford Linear Accelerator Center\\
Stanford University, Stanford, California 94309
\end{center}

\vspace{0.4cm}

\begin{abstract}
The short-distance expansion of the heavy-light currents $\bar
q\,\gamma^\mu Q$ and $\bar q\,\gamma^\mu\gamma_5\,Q$ is constructed to
order $1/m_Q$, and to next-to-leading order in
renor\-ma\-li\-zation-group improved perturbation theory. It is shown
that the $10\times 10$ anomalous dimension matrix, which describes the
scale dependence of the dimension-four effective current operators in
the heavy quark effective theory, is to a large extent determined by
the equations of motion, heavy quark symmetry, and reparameterization
invariance. The next-to-leading order expressions for the Wilson
coefficients at order $1/m_Q$ depend on only five unknown two-loop
anomalous dimensions, among them that of the chromo-magnetic operator.
\end{abstract}
\medskip
\centerline{(submitted to Physical Review D)}

\end{titlepage}

\section{Introduction}

Over the last few years, the heavy quark effective theory (HQET) has
been established as a convenient tool to analyze the properties of
hadrons containing a heavy quark
\cite{Eich1,Eich2,Geor,Mann,Falk,FGL,habil}. It provides a systematic
expansion of hadronic matrix elements around the limit
$\Lambda_{\rm QCD}/m_Q\to 0$, in which one or more heavy quark masses
tend to infinity and the effective low-energy Lagrangian of QCD
exhibits a spin-flavor symmetry under the substitution of heavy quarks
of different flavor or spin, but with the same velocity
\cite{Shur,Nuss,Volo,Isgu}. The existence of such a symmetry limit, and
the establishment of HQET to analyze the symmetry-breaking corrections
to it, helped to remove much of the model dependence from the
theoretical description of semileptonic weak decay processes of heavy
mesons or baryons.

The construction of a systematic $1/m_Q$ expansion of hadronic matrix
elements consists of two steps: One first constructs the $1/m_Q$
expansion of the weak currents and of the effective low-energy
Lagrangian using the operator product expansion and renormalization
group techniques. This part of the calculation involves short-distance
physics only and, in particular, is independent of the external states.
In a second step, hadronic matrix elements of the HQET operators are
parameterized by universal, $m_Q$-independent form factors. For the
case of transitions between two heavy quarks, the short-distance
expansion is known to order $1/m_Q$, and the corresponding Wilson
coefficients have been calculated to next-to-leading order in
perturbation theory. The necessary hadronic matrix elements for meson
and baryon decays have been investigated in detail up to order
$1/m_Q^2$. These results build a solid theoretical basis for a
model-independent extraction of the Cabibbo-Kobayashi-Maskawa matrix
element $V_{cb}$ from semileptonic $B$ decays, with a theoretical
uncertainty of only 5\% (for a review, see Ref.~\cite{habil} and
references therein).

The $1/m_Q$ expansion is less developed for heavy-to-light transitions,
which are, nevertheless, of great phenomenological importance: A
thorough understanding of exclusive decays such as $\bar B\to\ell\,
\bar\nu$ or $\bar B\to\pi\,\ell\,\bar\nu$ is necessary for a precise
determination of $V_{ub}$. For the most interesting processes of this
type, the constraints imposed by heavy quark symmetry have been
analyzed in the $m_Q\to\infty$ limit (see, e.g.,
Refs.~\cite{Volo,IWis,Wise,Gust,Dib}), but little is known about the
symmetry-breaking corrections. In particular, already at order $1/m_Q$
the short-distance expansion of the weak currents is only known in
leading logarithmic approximation \cite{FaGr}, and the hadronic matrix
elements appearing at that order have only been analyzed for the
simplest case, namely for heavy meson decay constants \cite{SR2}. In
this paper we address the first problem and reconsider the
short-distance expansion of heavy-light currents at order $1/m_Q$
beyond the leading logarithmic approximation. A detailed analysis of
the hadronic matrix elements for $\bar B\to\pi\,\ell\,\bar\nu$ decays
at order $1/m_Q$ will be presented elsewhere \cite{BtoPi}.

After a brief review of the formalism of HQET relevant to our work, we
calculate in Sec.~\ref{sec:3} the Wilson coefficients appearing in the
expansion of heavy-light currents at the one-loop level. This provides,
in particular, the matching corrections between QCD and HQET at
$\mu=m_Q$. In Sec.~\ref{sec:4}, we improve this calculation by using
standard renormalization group techniques to sum the leading and
next-to-leading logarithms $[\alpha_s\ln(m_Q/\mu)]^n$ and $\alpha_s
[\alpha_s\ln(m_Q/\mu)]^n$ to all orders in perturbation theory. This is
a complicated problem in that one has to deal with a set of ten
operators that mix under renormalization, with the technical
complication that the $10\times 10$ one-loop anomalous dimension matrix
cannot be fully diagonalized. For the complete next-to-leading order
solution of the renormalization group equation, one needs the anomalous
dimension matrix at two-loop order. Although we do not calculate this
matrix explicitly, we are able to infer much of its structure by
analyzing the constraints imposed by various symmetries of the
effective theory, as expressed in its Feynman rules, the equations of
motion, and an invariance under reparameterizations of the momentum
operator. From these considerations, we can determine the $10\times 10$
two-loop matrix up to only five unknown parameters. We then construct
the exact next-to-leading order solution for the Wilson coefficients in
terms of these parameters. In Sec.~\ref{sec:5}, we summarize our
results and give a sample application.

\section{Short-distance Expansion}
\label{sec:2}

In HQET, a heavy quark $Q$ bound inside a hadron moving at velocity $v$
is represented by a velocity-dependent field $h_v(x)$, which is related
to the conventional spinor field $Q(x)$ by \cite{Geor}
\begin{equation}
  h_v(x) = \exp(i m_Q\,v\!\cdot\! x)\,{1+\rlap/v\over 2}\,Q(x) \,.
\end{equation}
By means of the phase redefinition one removes the large part of the
heavy quark momentum from the new field. When the total momentum is
written as $p=m_Q\,v+k$, the field $h_v$ carries the residual momentum
$k$, which results from soft interactions of the heavy quark with light
degrees of freedom and is typically of order $\Lambda_{\rm QCD}$. The
operator $\frac{1}{2}(1+\rlap/v)$ projects out the heavy quark (rather
than antiquark) components of the spinor. The antiquark components are
integrated out to obtain the effective Lagrangian
\cite{Eich1,Eich2,Geor,Mann,FGL}
\begin{equation}\label{Leff}
   {\cal{L}}_{\rm eff} = \bar h_v\,i v\!\cdot\!D\,h_v
   + {1\over 2 m_Q}\,\Big[ O_{\rm kin}
   + C_{\rm mag}(\mu)\,O_{\rm mag} \Big] + {\cal{O}}(1/m_Q^2) \,,
\end{equation}
where $D^\mu = \partial^\mu - i g_s T_a A_a^\mu$ is the gauge-covariant
derivative. The operators appearing at
order $1/m_Q$ are
\begin{equation}\label{Omag}
   O_{\rm kin} = \bar h_v\,(i D)^2 h_v \,, \qquad
   O_{\rm mag} = {g_s\over 2}\,\bar h_v\,\sigma_{\mu\nu}
                 G^{\mu\nu} h_v \,.
\end{equation}
Here $G^{\mu\nu}$ is the gluon field strength tensor defined by
$[iD^\mu,iD^\nu]=i g_s G^{\mu\nu}$. In the hadron's rest frame, it is
readily seen that $O_{\rm kin}$ describes the kinetic energy resulting
from the residual motion of the heavy quark, whereas $O_{\rm mag}$
describes the chromo-magnetic coupling of the heavy quark spin to the
gluon field. One can show that, to all orders in perturbation theory,
the kinetic operator $O_{\rm kin}$ is not renormalized \cite{LuMa}. The
renormalization factor $C_{\rm mag}(\mu)$ of the chromo-magnetic
operator will be given later.

In order to construct a systematic $1/m_Q$ expansion, one works with
the eigenstates of the leading-order term in the effective Lagrangian
and treats the $1/m_Q$ corrections as a perturbation. The set of
operators that appear at order $1/m_Q$ can be reduced by using the
leading-order equation of motion $i v\!\cdot\!D\,h_v=0$, since the
physical matrix elements of operators which vanish by this equation are
at least of order $1/m_Q^2$. We have used this freedom to omit in
(\ref{Leff}) the operator $\bar h_v\,(i v\!\cdot\!D)^2 h_v$.

The goal of this paper is the construction of the short-distance
expansion for the heavy-light vector and axial vector currents
$V^\mu=\bar q\,\gamma^\mu Q$ and $A^\mu=\bar q\,\gamma^\mu\gamma_5\,Q$
in terms of operators of the effective theory, beyond the leading order
in $1/m_Q$. We shall discuss the case of the vector current in detail.
The general form of the expansion can be written as
\begin{equation}\label{Vexp}
   V^\mu \cong \sum_i C_i(\mu)\,J_i
   + {1\over 2 m_Q} \sum_j B_j(\mu)\,O_j
   + {1\over 2 m_Q} \sum_k A_k(\mu)\,T_k + {\cal{O}}(1/m_Q^2) \,.
\end{equation}
The symbol $\cong$ is used to indicate that this is an equation that
holds on the level of matrix elements. The operators $\{J_i\}$ form a
complete set of local dimension-three current operators with the same
quantum numbers as the original vector current in the full theory. In
HQET there are two such operators, namely
\begin{equation}
   J_1 = \bar q\,\gamma^\mu h_v \,, \qquad
   J_2 = \bar q\,v^\mu h_v \,.
\end{equation}
Similarly, $\{O_j\}$ are a complete set of local dimension-four
operators. It is convenient to use the background field method, which
ensures that there is no mixing between gauge-invariant and
gauge-dependent operators. Moreover, operators that vanish by the
equation of motion are irrelevant at order $1/m_Q$. It is thus
sufficient to consider gauge-invariant operators that do not vanish by
the equation of motion. A convenient basis of such operators is
\cite{AMM}:
\begin{equation}
   \begin{array}{ll}
   O_1 = \bar q\,\gamma^\mu\,i\,\rlap/\!D\,h_v \,, &\qquad
   O_4 = \bar q\,(-i v\!\cdot\!\overleftarrow{D})\,\gamma^\mu h_v \,,
   \\
   O_2 = \bar q\,v^\mu\,i\,\rlap/\!D\,h_v \,, &\qquad
   O_5 = \bar q\,(-i v\!\cdot\!\overleftarrow{D})\,v^\mu h_v \,,
   \\
   O_3 = \bar q\,i D^\mu h_v \,, &\qquad
   O_6 = \bar q\,(-i\overleftarrow{D^\mu})\,h_v \,.
   \end{array}
\end{equation}
For simplicity we consider here the limit where the light quark is
massless, $m_q=0$. Otherwise one would have to include two additional
operators $O_7=m_q\,J_1$ and $O_8=m_q\,J_2$.

In (\ref{Vexp}) we have also included nonlocal operators $T_k$ which
arise from an insertion of a $1/m_Q$ correction to the effective
Lagrangian into matrix elements of the leading-order currents:
\begin{eqnarray}
   T_1 &=& i\!\int\!{\rm d}y\,T\,\big\{J_1(0),O_{\rm kin}(y)\big\} \,,
    \nonumber\\
   T_2 &=& i\!\int\!{\rm d}y\,T\,\big\{J_2(0),O_{\rm kin}(y)\big\} \,,
    \nonumber\\
   T_3 &=& i\!\int\!{\rm d}y\,T\,\big\{J_1(0),O_{\rm mag}(y)\big\} \,,
    \nonumber\\
   T_4 &=& i\!\int\!{\rm d}y\,T\,\big\{J_2(0),O_{\rm mag}(y)\big\} \,.
\end{eqnarray}
The Wilson coefficients of these time-ordered products are simply the
products of the coefficients of their component operators, i.e.
\begin{eqnarray}\label{Aksol}
   A_1(\mu) &=& C_1(\mu) \,, \nonumber\\
   A_2(\mu) &=& C_2(\mu) \,, \nonumber\\
   A_3(\mu) &=& C_1(\mu)\,C_{\rm mag}(\mu) \,, \nonumber\\
   A_4(\mu) &=& C_2(\mu)\,C_{\rm mag}(\mu) \,.
\end{eqnarray}
Nevertheless, since these nonlocal operators can mix into the local
operators $O_j$ under renormalization (but not vice versa), it is
convenient to include them as parts of the effective currents
\cite{FaGr}.

Our goal is the calculation of the short-distance coefficients
$B_j(\mu)$ in (\ref{Vexp}) beyond the leading logarithmic
approximation. The coefficients $C_i(\mu)$ have been calculated at
next-to-leading order in Ref.~\cite{JiMu}. We shall perform the
calculation using dimensional regularization with modified minimal
subtraction ($\overline{\rm MS}$) and anticommuting $\gamma_5$. This
scheme has the advantage that, to all orders in $1/m_Q$, the operator
product expansion of the axial vector current $A^\mu$ can be simply
obtained from (\ref{Vexp}) by replacing $\bar q\to-\bar q\,\gamma_5$ in
the HQET operators. The Wilson coefficients remain unchanged
\cite{AMM}. The reason is that in any diagram the $\gamma_5$ from the
current can be moved outside next to the light quark spinor. For
$m_q=0$, this operation always leads to a minus sign. Hence it is
sufficient to consider the case of the vector current in detail.

Before we turn to explicit calculations, let us recall that there are
nontrivial relations between some of the coefficients $B_j(\mu)$ and
$C_i(\mu)$ imposed by a ``hidden'' symmetry of the effective theory,
namely its invariance under reparameterizations of the heavy quark
velocity and residual momentum which leave the total momentum
unchanged. One can show that, as a consequence, operators with a
covariant derivative acting on a heavy quark field must always appear
in certain combinations with lower-dimension operators \cite{LuMa}. For
the case at hand, there is a unique way in which the operators $O_1$,
$O_2$, $O_3$ can be combined with $J_1$, $J_2$ in a reparameterization
invariant form, namely
\begin{eqnarray}\label{rpiops}
   \bar q\,\gamma^\mu \bigg( 1 + {i\,\rlap/\!D\over 2 m_Q} \bigg) h_v
    + \ldots &=& J_1 + {O_1\over 2 m_Q} + \ldots \,, \nonumber\\
   \bar q\,\bigg( v^\mu + {i D^\mu\over m_Q} \bigg)
    \bigg( 1 + {i\,\rlap/\!D\over 2 m_Q} \bigg) h_v + \ldots
    &=& J_2 + {O_2 + 2 O_3\over 2 m_Q} + \ldots \,.
\end{eqnarray}
This implies that, to all orders in perturbation theory,
\begin{eqnarray}\label{rpirel}
   B_1(\mu) &=& C_1(\mu) \,, \nonumber\\
   B_2(\mu) &=& {1\over 2}\,B_3(\mu) = C_2(\mu) \,.
\end{eqnarray}
This is an important constraint, which has to be obeyed by any explicit
calculation.

\section{One-loop Matching}
\label{sec:3}

The Wilson coefficients in (\ref{Vexp}) are defined by requiring that
matrix elements of the vector current in the full theory agree, to any
order in $1/m_Q$, with matrix elements calculated in HQET. Order by
order in perturbation theory, these coefficients can be computed from a
comparison of matrix elements in the two theories. Since the effective
theory is constructed to reproduce correctly the low-energy behavior of
the full theory, this ``matching'' procedure is independent of any
long-distance physics such as infrared singularities, nonperturbative
effects, and the properties of the external states. There is thus a
freedom in the choice of the external states and the infrared
regularization scheme, which can be exploited to simplify the
calculations. We find it most convenient to perform the matching of QCD
onto HQET using on-shell external quark states and dimensional
regularization for both the ultraviolet and infrared singularities
encountered in the evaluation of loop diagrams.\footnote{The usefulness
of this scheme was pointed out by Eichten and Hill in the original
calculation of the renormalized effective Lagrangian at order $1/m_Q$
\cite{Eich2}.}
This scheme has the great advantage that all loop diagrams in the
effective theory vanish, since there is no mass scale other than the
renormalization point $\mu$. This means that matrix elements in HQET
are simply given by their tree-level expressions. We use momentum
assignments such that the incoming heavy quark has momentum
$p=m_Q\,v+k$ (with $2 v\cdot k+k^2/m_Q=0$), while the outgoing light
quark carries momentum $p'$ (with $p'^2=0$). Then, for instance,
\begin{eqnarray}
   \langle J_1\rangle &=& \bar u_q(p',s')\,\gamma^\mu\,u_h(v,s) \,,
    \nonumber\\
   \langle O_1\rangle &=& \bar u_q(p',s')\,\gamma^\mu\,\rlap/k\,
    u_h(v,s) \,,
\end{eqnarray}
etc., where $u_q(p',s')$ and $u_h(v,s)$ are on-shell spinors for a
massless quark $q$ and a heavy quark in HQET, respectively. They
satisfy $\rlap/\!p\,'\,u_q(p',s')=0$ and $\rlap/v\,u_h(v,s)=u_h(v,s)$.
To complete the matching calculation, one computes in the full theory
the vector current matrix element between on-shell quark states at
one-loop order. Taking into account that the relation between the heavy
quark spinors in QCD and in the effective theory is \cite{Mann}
\begin{equation}
   u_Q(p,s) = \bigg( 1 + {\rlap/k\over 2 m_Q} \bigg)\,u_h(v,s)
   + {\cal{O}}(1/m_Q^2) \,,
\end{equation}
we find at one-loop order
\begin{eqnarray}
   \langle V^\mu\rangle_{\rm QCD} &=&
    \bigg\langle J_1 + {O_1\over 2 m_Q}\bigg\rangle
    + {\alpha_s\over 3\pi}\,{\Gamma(2-D/2)\over D-3}\,
    \bigg({m_Q^2\over 4\pi\mu^2}\bigg)^{D/2-2} \nonumber\\
   &&\times \Bigg\{ {D-7\over 2}\,\bigg\langle J_1
    + {O_1\over 2 m_Q}\bigg\rangle
    - (D-4)\,\bigg\langle J_2 + {O_2 + 2 O_3\over 2 m_Q} \bigg\rangle
    \nonumber\\
   &&\quad + 2\,{D^2-9D+23\over D-5}\,\bigg\langle {O_4\over 2 m_Q}
    \bigg\rangle - 2(D-6)\,\bigg\langle {O_5\over 2 m_Q}\bigg\rangle
    + 2\,\bigg\langle {O_6\over 2 m_Q}\bigg\rangle \Bigg\}
    \nonumber\\
   &&+ {\cal{O}}(1/m_Q^2) \,,
\end{eqnarray}
where $D$ is the space-time dimension. The matrix elements on the
right-hand side are evaluated in the effective theory. Note that the
operators $J_1$ and $O_1$, as well as $J_2$, $O_2$ and $O_3$, appear
precisely in the combinations required by reparameterization
invariance, cf.~(\ref{rpiops}). Hence, our explicit one-loop
calculation is in accordance with the general relations (\ref{rpirel}).

Next we expand the above expressions around $D=4$ and subtract the
poles in
\begin{equation}
   {1\over\epsilon} = {2\over 4-D} - \gamma_E + \ln 4\pi \,,
\end{equation}
corresponding to an $\overline{\rm MS}$ renormalization of the
operators in the effective theory, to obtain the following one-loop
expressions for the Wilson coefficients:
\begin{eqnarray}\label{Bjres}
   C_1(\mu) &=& B_1(\mu) = 1 + {\alpha_s\over\pi}\,\bigg(
    \ln{m_Q\over\mu} - {4\over 3} \bigg) \,, \nonumber\\
   C_2(\mu) &=& B_2(\mu) = {1\over 2}\,B_3(\mu)
    = {2\alpha_s\over 3\pi} \,, \phantom{ \bigg( } \nonumber\\
   B_4(\mu) &=& \phantom{-} {4\alpha_s\over 3\pi}\,
    \bigg( 3\ln{m_Q\over\mu} - 1 \bigg) \,, \nonumber\\
   B_5(\mu) &=& - {4\alpha_s\over 3\pi}\,
    \bigg( 2\ln{m_Q\over\mu} - 3 \bigg) \,, \nonumber\\
   B_6(\mu) &=& - {4\alpha_s\over 3\pi}\,
    \bigg( \ln{m_Q\over\mu} - 1 \bigg) \,.
\end{eqnarray}
The one-loop expressions for $C_i(\mu)$ have been first obtained in
Ref.~\cite{JiMu}. Our expressions for $B_j(\mu)$ are new.

\section{Renormalization Group Improvement}
\label{sec:4}

\subsection{Introduction}

For $\mu\ll m_Q$, the calculation presented above becomes
unsatisfactory, since the scale in the running coupling constant cannot
be determined at one-loop order. Yet $\alpha_s(\mu)$ and
$\alpha_s(m_Q)$ may differ substantially, and one would thus like to
resolve the scale ambiguity problem by going beyond the leading order
in perturbation theory. Furthermore, the perturbative expansion of the
Wilson coefficients is known to contain large logarithms of the type
$[\alpha_s\ln(m_Q/\mu)]^n$, which one should sum to all orders. Both
goals can be achieved by using the renormalization group to improve the
one-loop results derived in the previous section.

Let us introduce a compact matrix notation where we collect the
renormalized operators $O_j$ and $T_k$, as well as the coefficient
functions $B_j(\mu)$ and $A_k(\mu)$, into 10-component vectors
\begin{eqnarray}
   \vec O &=& (O_1,\ldots,O_6,T_1,\ldots,T_4) \,, \nonumber\\
   \vec B &=& (B_1,\ldots,B_6,A_1,\ldots,A_4) \,.
\end{eqnarray}
The Wilson coefficients obey a renormalization group equation
\begin{equation}
   \bigg(\mu {{\rm d}\over{\rm d}\mu} - \hat\gamma^t \bigg)\,
   \vec B(\mu) = 0 \,,
\end{equation}
where $\hat\gamma^t$ is the transposed $10\times 10$ anomalous
dimension matrix. It is defined in terms of the matrix $\hat Z$ that
relates the ``bare'' operators to the renormalized ones: $\vec O_{\rm
bare}=\hat Z\,\vec O$. In the $\overline{\rm MS}$ scheme, $\hat Z$
obeys an expansion
\begin{equation}
   \hat Z = \hat 1 + \sum_k {1\over\epsilon^k}\,\hat Z_k(g_s) \,,
\end{equation}
with coefficients $\hat Z_k(g_s)$ that depend on the renormalized
coupling constant $g_s$. Given this expansion, one can show that
\cite{Gros}
\begin{equation}
   \hat\gamma = - g_s\,{\partial\over\partial g_s}\,\hat Z_1(g_s) \,,
\end{equation}
meaning that the anomalous dimension matrix can be computed from the
$1/\epsilon$ poles in $\hat Z$.

The formal solution of the renormalization group equation reads
\begin{equation}\label{RGEsol}
   \vec B(\mu) = \hat U(\mu,m_Q)\,\vec B(m_Q) \,,
\end{equation}
with the evolution matrix \cite{Bur1,Bur2}
\begin{equation}\label{Uevol}
   \hat U(\mu,m_Q) = T_g\,\exp\!\int
   \limits_{\displaystyle g_s(m_Q)}^{\displaystyle g_s(\mu)}\!
   {\rm d}g\,{\hat\gamma^t(g)\over\beta(g)} \,.
\end{equation}
The $\beta$-function $\beta(g_s)=\mu{\rm d}g_s/{\rm d}\mu$ describes
the scale-dependence of the renormalized coupling constant in QCD. The
symbol ``$T_g$'' means an ordering in the coupling constant such that
the couplings increase from right to left (for $\mu<m_Q$). This is
necessary since, in general, the anomalous dimension matrices at
different values of $g$ do not commute: $[\hat\gamma(g_1),
\hat\gamma(g_2)]\ne 0$. Eq.~(\ref{Uevol}) can be solved perturbatively
by expanding the $\beta$-function and the anomalous dimension matrix in
the renormalized coupling constant:
\begin{eqnarray}\label{bcexp}
   \beta(g) &=& - \beta_0\,{g^3\over 16\pi^2}
    - \beta_1\,{g^5\over(16\pi^2)^2} + \ldots \,, \nonumber\\
   \hat\gamma(g) &=& \phantom{-} \hat\gamma_0\,{g^2\over 16\pi^2}
    + \hat\gamma_1\,\bigg({g^2\over 16\pi^2}\bigg)^2 + \ldots \,.
\end{eqnarray}
Here
\begin{equation}
   \beta_0 = 11 - {2\over 3}\,n_f \,, \qquad
   \beta_1 = 102 - {38\over 3}\,n_f \,,
\end{equation}
where $n_f$ denotes the number of light quark flavors with mass below
$m_Q$. At next-to-leading order in renormalization-group improved
perturbation theory, the evolution matrix can be written in the form
\cite{Bur2}
\begin{equation}
   \hat U(\mu,m_Q) = \bigg\{ 1 - {\alpha_s(\mu)\over 4\pi}
   \hat S \bigg\} \,\hat U_0(\mu,m_Q)\,
   \bigg\{ 1 + {\alpha_s(m_Q)\over 4\pi} \hat S \bigg\} + \ldots \,,
\end{equation}
where
\begin{equation}\label{ULLA}
   \hat U_0(\mu,m_Q) = \exp\bigg\{ {\hat\gamma_0^t\over 2\beta_0}\,
   \ln{\alpha_s(m_Q)\over\alpha_s(\mu)} \bigg\}
\end{equation}
describes the evolution in leading logarithmic approximation, and
$\hat S$ contains the next-to-leading corrections. This matrix is
defined by the algebraic equation
\begin{equation}\label{Seq}
   \hat S + {1\over 2\beta_0}\,\big[ \hat\gamma_0^t,\hat S \big]
   = {1\over 2\beta_0}\,\hat\gamma_1^t
   - {\beta_1\over 2\beta_0^2}\,\hat\gamma_0^t \,.
\end{equation}
Finally, we expand the initial values of the coefficient functions at
$\mu=m_Q$ as
\begin{equation}
   \vec B(m_Q) = \vec B_0 + {\alpha_s(m_Q)\over 4\pi}\,\vec B_1
   + \ldots \,.
\end{equation}
{}From our one-loop results in (\ref{Bjres}) we obtain
\begin{eqnarray}\label{B01}
   \vec B_0 &=& (1,0,0,0,0,0,1,0,1,0) \,, \phantom{ \bigg( }
    \nonumber\\
   \vec B_1 &=& {2\over 3}\,(-8,4,8,-8,24,8,-8,4,5,4) \,,
\end{eqnarray}
where we have used that \cite{Eich2}
\begin{equation}
   C_{\rm mag}(m_Q) = 1 + {26\over 3}\,{\alpha_s(m_Q)\over 4\pi} \,.
\end{equation}
Putting everything together, we obtain from (\ref{RGEsol})
\begin{eqnarray}\label{NLOsol}
   \vec B(\mu) &=& \hat U_0(\mu,m_Q)\,\bigg[ \vec B_0
    + {\alpha_s(m_Q)\over 4\pi}\,\Big( \vec B_1 + \hat S\,\vec B_0
    \Big) \bigg] \nonumber\\
   &&\mbox{}- {\alpha_s(\mu)\over 4\pi}\,\hat S\,
    \hat U_0(\mu,m_Q)\,\vec B_0 + \ldots \,.
\end{eqnarray}
What remains to be calculated are the matrices $\hat U_0$ and $\hat S$.
To this end, we need to know the $10\times 10$ anomalous dimension
matrix $\hat\gamma$ at two-loop order. Note that the leading-log
evolution matrix, as well as the coefficient $(\vec B_1+\hat S\,
\vec B_0)$ that multiplies $\alpha_s(m_Q)$, are independent of the
renormalization scheme \cite{Bur1}. The matrix $\hat S$ that multiplies
$\alpha_s(\mu)$, on the other hand, is scheme-dependent. Only when the
$\mu$-dependent terms in $\vec B(\mu)$ are combined with the
$\mu$-dependent matrix elements of the renormalized HQET operators does
one obtain a renormalization-scheme invariant result.

\subsection{Anomalous dimension matrix}

Let us now analyze the structure of the anomalous dimension matrix in
more detail. We will see that, to a large extent, the texture of this
matrix can be determined without an explicit calculation. From
(\ref{Aksol}) and (\ref{rpirel}) we know that the operators $O_1$,
$O_2$, $O_3$ and $T_k$ renormalize multiplicatively. They can mix into
$O_4$, $O_5$, $O_6$, but not vice versa. Therefore, the anomalous
dimension matrix must be of the form
\begin{equation}\label{texture}
   \hat\gamma = \left(\begin{array}{ccc}
       \hat\gamma_{\rm hl} ~&~ \hat\gamma_A ~&~ 0 \\
       0 ~&~ \hat\gamma_B ~&~ 0 \\
       0 ~&~ \hat\gamma_C ~&~ \hat\gamma_D
       \end{array}\right) ,
\end{equation}
with diagonal matrices
\begin{eqnarray}
   \hat\gamma_{\rm hl} &=& {\rm diag}\Big( \gamma^{\rm hl},
    \gamma^{\rm hl}, \gamma^{\rm hl} \Big) \,, \nonumber\\
   \hat\gamma_D &=& {\rm diag}\Big( \gamma^{\rm hl}, \gamma^{\rm hl},
    \gamma^{\rm hl} + \gamma^{\rm mag}, \gamma^{\rm hl}
    + \gamma^{\rm mag} \Big) \,.
\end{eqnarray}
Here $\gamma^{\rm hl}$ is the universal anomalous dimension of
dimension-three heavy-light current operators in HQET, and
$\gamma^ {\rm mag}$ is the anomalous dimension of the chromo-magnetic
operator in (\ref{Omag}). The form of $\hat\gamma_{\rm hl}$ is a
consequence of reparameterization invariance [cf.~(\ref{rpirel})],
whereas the structure of $\hat\gamma_D$ follows from (\ref{Aksol}). The
one-loop coefficients of the anomalous dimensions are \cite{Volo,FGL}
\begin{equation}
   \gamma_0^{\rm hl} = -4 \,, \qquad \gamma_0^{\rm mag} = 6 \,.
\end{equation}
The two-loop coefficient $\gamma_1^{\rm hl} = - \frac{254}{9}
- \frac{56}{27}\pi^2 + \frac{20}{9} n_f$ has been calculated in
Refs.~\cite{JiMu,BrGr}, whereas $\gamma_1^{\rm mag}$ is not yet known.

The $3\times 3$ submatrix $\hat\gamma_B$ in (\ref{texture}) may be
constructed by noting that the equation of motion $i v\!\cdot\!D\,h_v
=0$ can be used to rewrite
\begin{eqnarray}
   O_4 &=& - i v\cdot\partial\,J_1 \,, \nonumber\\
   O_5 &=& - i v\cdot\partial\,J_2 \,, \nonumber\\
   O_6 - O_3 &=& - i \partial^\mu\,\big[ \bar q\,h_v \big] \,.
\end{eqnarray}
The total derivatives of the currents on the right-hand side
renormalize multiplicatively and in the same way as the dimension-three
operators $J_i$. The additional power of the external momentum carried
by the current does not affect the divergences of loop diagrams. It
follows that
\begin{equation}
   \big(\hat\gamma_B\big)_{ij} = \gamma^{\rm hl}\,\delta_{ij}
   + \delta_{i3}\,\big(\hat\gamma_A\big)_{3j} \,.
\end{equation}

Let us then focus on the $3\times 3$ submatrix $\hat\gamma_A$, which
describes the mixing of $O_1$, $O_2$, $O_3$ into $O_4$, $O_5$, $O_6$.
To compute $\hat\gamma_A$, one has to calculate in the effective theory
the ultraviolet divergent $1/\epsilon$ poles of the $p'$-dependent
terms in the matrix element of the generic operator $\bar q\,\Gamma\,i
D_\alpha h_v$, where $p'$ denotes the external momentum of the light
quark. At the end of the calculation, one substitutes $\Gamma =
\gamma^\mu\gamma^\alpha$, $v^\mu\gamma^\alpha$, $g^{\mu\alpha}$ for
$O_1$, $O_2$, $O_3$, respectively. The structure of these terms is
strongly constrained by the form of the Feynman rules of the effective
theory, and by the equations of motion. In HQET, a heavy quark couples
to gluons proportional to its velocity \cite{Geor,Falk}, and hence
there do not appear any Dirac matrices on the right-hand side of
$\Gamma$. Furthermore, since we consider the limit of a massless light
quark, there can only be an even number of $\gamma$-matrices on the
left-hand side of $\Gamma$, since vertices and propagators come in
pairs. Moreover, the equation of motion for the heavy quark, $i
v\!\cdot\!D\,h_v=0$, requires that the matrix elements must vanish upon
contraction with $v^\alpha$. Likewise, the equation of motion for the
light quark, $i\,\rlap/\!D\,q=0$, implies that the $p'$-dependent terms
must vanish when one substitutes $\Gamma=\gamma^\alpha\Gamma'$. Using
these constraints, we find that, to all orders in perturbation theory,
the ultraviolet divergent pole in the matrix element must be of the form
\begin{eqnarray}
   \langle\,\bar q\,\Gamma\,i D_\alpha h_v\,\rangle\Big|_{\rm pole}
   &=& {f_1(g_s)\over 3\epsilon}\,\bar u_q(p')\,\Big[
    3 p'_\alpha - 2 v\cdot p'\,v_\alpha - v\cdot p'\,\gamma_\alpha\,
    \rlap/v \Big]\,\Gamma\,u_h(v) \nonumber\\
   \phantom{ \bigg( }
   &&\mbox{}+ \hbox{\rm $p'$-independent terms,}
\end{eqnarray}
where the function $f_1(g_s)$ has a perturbative expansion in powers of
the coupling constant. From this, it is straightforward to derive that
\begin{equation}
   \hat\gamma_A = \gamma^a\,
   \left(\begin{array}{rrr}
         -\frac{2}{3} ~& -\frac{4}{3} ~&~ 2 \\
         0 ~& 0 ~&~ 0 \\
         -\frac{1}{3} ~& -\frac{2}{3} ~&~ 1
   \end{array}\right) ,
\end{equation}
where
\begin{equation}
   \gamma^a = - g_s\,{\partial f_1\over\partial g_s} \,.
\end{equation}

We can use similar arguments to impose some restrictions on the
$4\times 3$ submatrix $\hat\gamma_D$, which describes the mixing of the
nonlocal operators $T_k$ into $O_4$, $O_5$, $O_6$. The Feynman rules of
HQET imply that
\begin{eqnarray}
   \langle\,T_{1,2}\,\rangle\Big|_{\rm pole}
   &=& {f_2(g_s)\over\epsilon}\,v\cdot p'\,\bar u_q(p')\,\Gamma\,
    u_h(v) + \hbox{\rm $p'$-independent terms,} \nonumber\\
   \phantom{ \Bigg( }
   \langle\,T_{3,4}\,\rangle\Big|_{\rm pole}
   &=& -{1\over 4\epsilon}\,\bar u_q(p') \Big[ f_3(g_s)\,
    v\cdot p'\,\sigma_{\alpha\beta} + i f_4(g_s)\,\gamma_\alpha
    p'_\beta\,\rlap/v \Big]\,\Gamma\,(1+\rlap/v)
    \sigma^{\alpha\beta} u_h(v) \nonumber\\
   \phantom{ \bigg( }
   &&\mbox{}+ \hbox{\rm $p'$-independent terms,}
\end{eqnarray}
where $\Gamma=\gamma^\mu$ for $T_1$ and $T_3$, whereas $\Gamma=v^\mu$
for $T_2$ and $T_4$. It follows that
\begin{equation}
   \hat\gamma_D =
   \left(\begin{array}{ccc} \gamma^b ~~& 0 &~~ 0 \\
         0 ~~& \gamma^b &~~ 0 \\
         \gamma^c ~~& -4\gamma^c-2\gamma^d &~~ \gamma^d \\
         0 ~~& -3\gamma^c-\gamma^d &~~ 0
    \end{array}\right) ,
\end{equation}
where
\begin{equation}
   \gamma^b = - g_s\,{\partial f_2\over\partial g_s} \,, \qquad
   \gamma^c = - g_s\,{\partial f_3\over\partial g_s} \,, \qquad
   \gamma^d = - g_s\,{\partial f_4\over\partial g_s} \,.
\end{equation}

We have calculated the anomalous dimension matrix $\hat\gamma$ at the
one-loop level by computing the ultraviolet divergences of the bare
operators $\vec O_{\rm bare}$. Our results are in agreement with the
general relations derived above. Expanding the anomalous dimensions
$\gamma^a,\ldots,\gamma^d$ as in (\ref{bcexp}), we find the one-loop
coefficients
\begin{equation}
   \gamma_0^a = 4 \,, \qquad \gamma_0^b = -{32\over 3} \,, \qquad
   \gamma_0^c = \gamma_0^d = -{8\over 3} \,.
\end{equation}
We note that the first row in $\hat\gamma_A$, as well as the first and
third rows in $\hat\gamma_C$, can be derived from a previous one-loop
calculation by Falk and Grinstein \cite{FaGr}. We confirm their results.

\subsection{Leading-log solution}

Given the one-loop anomalous dimension matrix, one can calculate the
evolution matrix in leading logarithmic approximation from
(\ref{ULLA}). The evaluation of this equation would be straightforward
if there were a matrix $\hat V$ such that $\hat V^{-1}\hat\gamma_0^t\,
\hat V$ were diagonal \cite{Bur2}. However, such a matrix does not
exist in the present case. The best one can achieve is to construct a
matrix $\hat W$ which brings $\hat\gamma_0^t$ into Jordan form:
\begin{equation}
   \hat W^{-1} \hat\gamma_0^t\,\hat W = \hat\gamma_J \,.
\end{equation}
A matrix which accomplishes this is:
\begin{equation}\label{What}
   \hat W = \left(
   \begin{array}{cccccccccc}
    0 & 0 & 0 & 0 & 0 & 0 & 0 & 0 & 0 & 1 \\
    0 & 0 & 0 & 0 & 0 & 0 & 1 & 0 & -1 & 0 \\
    0 & 0 & 0 & 0 & 0 & 0 & 0 & 1 & 0 & 0 \\
    \phantom{\Big.}
    -{1\over 3} & 0 & 0 & 0 & -{4\over 27} & -{32\over 3} &
     {16\over 27} & {1\over 3} & -{64\over 27} & {2\over 3} \\
    \phantom{\bigg.}
    -{2\over 3} & {16\over 9} & -{32\over 3} & -{16\over 9} &
     {8\over 9} & {128\over 9} & 0 & {26\over 3} & -{32\over 27} &
     {52\over 3} \\
    \phantom{\bigg.}
    1 & 0 & 0 & 0 & -{4\over 3} & 0 & 0 & -1 & 0 & -2 \\
    0 & 0 & 0 & 0 & 0 & 0 & 1 & 0 & 0 & 0 \\
    0 & 0 & 0 & 1 & 0 & 0 & -{4\over 3} & 0 & 0 & 0 \\
    0 & 0 & 0 & 0 & 1 & 0 & 0 & 0 & 0 & 0 \\
    0 & 1 & 0 & 0 & -{4\over 3} & 0 & 0 & 0 & 0 & 0
   \end{array}
   \right)
\end{equation}
A Jordan matrix is convenient enough for an exponentiation in closed
form. In our case, we only need that
\begin{equation}
   \exp\left( \begin{array}{cc} a &~ b\\ 0 &~ a \end{array}
   \right) = \exp(a) \left(
   \begin{array}{cc} 1 &~ b\\ 0 &~ 1 \end{array} \right) \,.
\end{equation}
It is then straightforward to compute the evolution matrix from
\begin{equation}\label{Wdef}
   \hat U_0(\mu,m_Q) = \hat W\,\exp\bigg\{
   {\hat\gamma_J\over 2\beta_0}\,\ln{\alpha_s(m_Q)\over\alpha_s(\mu)}
   \bigg\}\,\hat W^{-1} \,.
\end{equation}
Multiplying $\hat U_0$ with the tree-level matching condition as
encoded in $\vec B_0$ in (\ref{B01}), we recover the leading-log
results of Refs.~\cite{FaGr,AMM}:
\begin{eqnarray}\label{BjLLA}
   \phantom{ \bigg( }
   C_1(\mu) &=& B_1(\mu) = x^{2/\beta_0} \,, \nonumber\\
   \phantom{ \bigg( }
   C_2(\mu) &=& B_2(\mu) = B_3(\mu) = 0 \,, \nonumber\\
   B_4(\mu) &=& \phantom{-} {34\over 27}\,x^{2/\beta_0}
    - {4\over 27}\,x^{-1/\beta_0} - {10\over 9}
    + {16\over 3\beta_0}\,x^{2/\beta_0} \ln x \,, \nonumber\\
   B_5(\mu) &=& - {28\over 27}\,x^{2/\beta_0}
    + {88\over 27}\,x^{-1/\beta_0} - {20\over 9} \,, \nonumber\\
   B_6(\mu) &=& - 2\,x^{2/\beta_0} - {4\over 3}\,x^{-1/\beta_0}
    + {10\over 3} \,,
\end{eqnarray}
with
\begin{equation}
   x = {\alpha_s(\mu)\over\alpha_s(m_Q)} \,.
\end{equation}
For $m_Q\approx\mu$, these expressions can be expanded using
\begin{equation}
   x \approx 1 + 2\beta_0\,{\alpha_s\over 4\pi}\,\ln{m_Q\over\mu} \,,
\end{equation}
and one readily recovers the logarithmic terms in the one-loop
coefficients given in (\ref{Bjres}).

At this point, it is worthwhile to compare the the leading logarithmic
approximation with the one-loop results. For the purpose of
illustration, we use $\alpha_s(m_Q)=0.20$ and $\alpha_s(\mu)=0.36$
(corresponding to $m_Q\approx m_b$ and $\mu\approx 1$ GeV), as well as
$n_f=4$. From (\ref{BjLLA}), we then obtain: $B_1^{\rm LL}=1.15$,
$B_2^{\rm LL}= B_3^{\rm LL}=0$, $B_4^{\rm LL}=0.63$,
$B_5^{\rm LL}=-0.38$, $B_6^{\rm LL}=-0.21$. In the one-loop expressions
(\ref{Bjres}), we use the average value $\alpha_s=0.28$, as well as
$m_Q/\mu=4.8$. This gives: $B_1^{(1)}=1.02$, $B_2^{(1)}=0.06$,
$B_3^{(1)}=0.12$, $B_4^{(1)}=0.44$, $B_5^{(1)}=-0.02$,
$B_6^{(1)}=-0.07$. Obviously, there are significant differences between
these two approximation schemes, which can only be resolved by going
beyond the leading order.

\subsection{Next-to-leading order solution}

Let us then discuss the construction of the complete next-to-leading
order solution of the renormalization group equation. According to
(\ref{NLOsol}), we need to construct the matrix $\hat S$ that satisfies
the algebraic equation (\ref{Seq}). To this end, it is convenient to
define $\hat T=\hat W^{-1} \hat S\,\hat W$, where $\hat W$ has been
given in (\ref{What}). The algebraic equation which determines $\hat T$
reads
\begin{equation}
   \hat T + {1\over 2\beta_0}\,\big[ \hat\gamma_J,\hat T \big]
   = {1\over 2\beta_0}\,\hat W^{-1} \hat\gamma_1^t\,\hat W
   - {\beta_1\over 2\beta_0^2}\,\hat\gamma_J \,.
\end{equation}
We have solved this equation be modifying the Jordan matrix,
$\hat\gamma_J\to\hat\gamma_J(\eta)$, such that $\hat\gamma_J(0)\equiv
\hat\gamma_J$, but $\hat\gamma_J(\eta)$ can be diagonalized for
$\eta\ne 0$. It is then straightforward to construct a matrix
$\hat T(\eta)$ which satisfies the above equation with $\hat\gamma_J$
replaced by $\hat\gamma_J(\eta)$ \cite{Bur2}. The desired matrix $\hat
T$ is obtained by taking the limit $\eta\to 0$. We find that most of
the entries in $\hat T$ vanish. Among the nonvanishing components
$T_{ij}$ are:
\begin{eqnarray}
   T_{ii} &=& S_{\rm hl} + S_{\rm mag} \,;\quad i=2,5, \nonumber\\
   T_{ii} &=& S_{\rm hl} \,;\quad i=3,4,6,7,8,9,10,
\end{eqnarray}
where
\begin{equation}
   S_{\rm hl} = {\gamma_1^{\rm hl}\over 2\beta_0}
    - {\beta_1\gamma_0^{\rm hl}\over 2\beta_0^2} \,, \qquad
   S_{\rm mag} = {\gamma_1^{\rm mag}\over 2\beta_0}
    - {\beta_1\gamma_0^{\rm mag}\over 2\beta_0^2} \,.
\end{equation}
These are nothing but the next-to-leading corrections to the Wilson
coefficients of the dimension-three heavy-light currents and of the
chromo-magnetic operator, which are given by \cite{JiMu,Eich2,FGL}:
\begin{eqnarray}\label{Cisolu}
   C_1(\mu) &=& x^{2/\beta_0}\,\bigg\{ 1
    + {\alpha_s(m_Q)-\alpha_s(\mu)\over 4\pi}\,S_{\rm hl}
    - {4\over 3}\,{\alpha_s(m_Q)\over\pi} \bigg\} \,, \nonumber\\
   C_2(\mu) &=& {2\over 3}\,x^{2/\beta_0}\,{\alpha_s(m_Q)\over\pi}
    \,, \nonumber\\
   C_{\rm mag}(\mu) &=& x^{-3/\beta_0}\,\bigg\{ 1
    + {\alpha_s(m_Q)-\alpha_s(\mu)\over 4\pi}\,S_{\rm mag}
    + {13\over 6}\,{\alpha_s(m_Q)\over\pi} \bigg\} \,.
\end{eqnarray}
As previously, $x=\alpha_s(\mu)/\alpha_s(m_Q)$. The remaining
nonvanishing components of $\hat T$ are:
\begin{eqnarray}
   T_{11} &=& {\gamma_1^{\rm hl} + \gamma_1^a\over 2\beta_0} \,,
    \nonumber\\
   T_{15} &=& {4\gamma_1^{\rm mag} - 4\gamma_1^a + 3\gamma_1^d\over
    6(\beta_0-1)} \,, \nonumber\\
   T_{32} &=& - {9\over 4}\,T_{35} = -3\,T_{65}
    = {16\gamma_1^{\rm mag} + 27\gamma_1^c + 9\gamma_1^d\over
    192(\beta_0-3)}  \,, \nonumber\\
   T_{34} &=& T_{67} = - {3\gamma_1^b\over 64\beta_0}
    - {\beta_1\over 2\beta_0^2} \,.
\end{eqnarray}

Given these results, one can compute the Wilson coefficients from
(\ref{NLOsol}). For the coefficients $A_k(\mu)$ of the nonlocal
operators $T_k$, and for the first three of the coefficients
$B_j(\mu)$, we confirm the general relations (\ref{Aksol}) and
(\ref{rpirel}) at next-to-leading order. The exact next-to-leading
order results for the remaining coefficients are rather complicated. We
present them in the form of three independent combinations:
\begin{eqnarray}\label{magic}
   4 B_4 &+& B_5 + 2 B_6 = {128\over 3}\,C_1\,\bigg\{
    {1\over 2\beta_0}\,\ln x - T_{34}\,
    {\alpha_s(m_Q)-\alpha_s(\mu)\over 4\pi} \bigg\} \nonumber\\
   &&\phantom{ B_5 + 2 B_6 = }\mbox{}
    + {2\over 9}\,C_2\,\bigg\{ 1 + 8\,C_{\rm mag}
    + {24\over\beta_0}\,\ln x \bigg\} \,, \nonumber\\
   3 B_5 &+& 2 B_6 = - {64\over 9}\,C_1\,\bigg\{ \bigg[ 1
    + 6\,T_{32}\,{\alpha_s(m_Q)\over 4\pi} \bigg] - C_{\rm mag}\,\bigg[
    1 + 6\,T_{32}\,{\alpha_s(\mu)\over 4\pi} \bigg] \bigg\} \nonumber\\
   &&\phantom{ 2 B_6 = }\mbox{}
    + {16\over 3}\,C_2\,\bigg\{ C_{\rm mag} - {29\over 24}
    + {3\over\beta_0}\,\ln x \bigg\} \,, \nonumber\\
   B_6 &=& - 2\,C_1 - 2\,C_2 - {4\over 3}\,C_1\,C_{\rm mag}\,\bigg\{
    1 + {3\over 4}\,T_{15}\,{\alpha_s(\mu)\over 4\pi} \bigg\}
    + T_{15}\,{\alpha_s(m_Q)\over 4\pi} \nonumber\\
   &&\mbox{}+ {10\over 3}\,\bigg\{ 1 + {\alpha_s(m_Q)\over 3\pi}
    + T_{11}\,{\alpha_s(m_Q)-\alpha_s(\mu)\over 4\pi} \bigg\} \,,
\end{eqnarray}
where we have omitted the $\mu$-dependence of the coefficient functions
for simplicity. The coefficients $C_i(\mu)$ have been given in
(\ref{Cisolu}).

\section{Summary and an Application}
\label{sec:5}

We have presented a detailed analysis of the short-distance expansion
of the heavy-light currents $\bar q\,\gamma^\mu Q$ and $\bar q\,
\gamma^\mu\gamma_5\,Q$ to order $1/m_Q$ in the heavy quark expansion,
and to next-to-leading order in renormalization-group improved
perturbation theory. Our main result is that the $10\times 10$
anomalous dimension matrix $\hat\gamma$, which describes the scale
dependence of the dimension-four effective current operators in the
heavy quark effective theory, can be determined to a large extent from
symmetry considerations. By evaluating the constraints that arise from
reparameterization invariance, the equations of motion, and heavy quark
symmetry, we have shown that, to all orders in perturbation theory,
$\hat\gamma$ can be expressed in terms of the universal anomalous
dimension of the leading order (dimension-three) currents, the
anomalous dimension of the chromo-magnetic operator, and four functions
$\gamma^i(g_s)$ of the coupling constant. We have calculated these
functions to one-loop order, and have treated their two-loop
coefficients, as well as the yet unknown two-loop anomalous dimension
of the chromo-magnetic operator, as free parameters in the exact
next-to-leading order solution of the renormalization group equation.
Our final expressions for the Wilson coefficients appearing at order
$1/m_Q$ in the expansion of the currents are given in (\ref{Aksol}),
(\ref{rpirel}), and (\ref{magic}).

We believe that the fact that our results depend on only five unknown
two-loop anomalous dimensions (as compared to the \`a priori 100
unknown entries in the two-loop anomalous dimension matrix) makes it
feasible to obtain the complete next-to-leading order solution for the
Wilson coefficients in the near future. This would be desirable since,
as we have pointed out, there are substantial numerical differences in
the results obtained in leading logarithmic approximation as compared
to those obtained from a one-loop calculation, which we have presented
in Sec.~\ref{sec:3}. These differences can only be resolved at
next-to-leading order.

The short-distance expansion of heavy-light currents considered in this
paper plays an important role in applications of the heavy quark
effective theory to decay processes such as $\bar B\to\ell\,\bar\nu$ or
$\bar B\to X\,\ell\, \bar\nu$, where $X$ is a light meson ($X=\pi,
\rho$, etc.). As an example, we briefly discuss the case of meson decay
constants, following the analysis of Ref.~\cite{SR2}. To order $1/m_Q$
in the heavy quark expansion, the decay constants of pseudoscalar and
vector mesons can be written as
\begin{eqnarray}\label{fPfV}
   f_M\sqrt{m_M} &=& \bigg[ C_1(\mu) + {(1+d_M)\over 4}\,C_2(\mu)
    \bigg]\,F(\mu) \\
   &\times& \bigg\{ 1 + {1\over m_Q} \bigg[ G_1(\mu)
    - {\bar\Lambda\over 6}\,b(\mu) \bigg] + {2 d_M\over m_Q} \bigg[
    C_{\rm mag}(\mu)\,G_2(\mu) - {\bar\Lambda\over 12}\,B(\mu) \bigg]
    \bigg\} \,, \nonumber
\end{eqnarray}
where $d_M=3$ if $M$ is a pseudoscalar meson, and $d_M=-1$ if $M$ is a
vector meson. $F(\mu)$, $G_i(\mu)$, and $\bar\Lambda$ are
$m_Q$-independent low-energy parameters of the effective theory. The
coefficients $b(\mu)$ and $B(\mu)$ contain combinations of the Wilson
coefficients $B_j(\mu)$, which we have calculated in this paper. One
obtains \cite{SR2}
\begin{eqnarray}
   b &=& {3\over 4(C_1+C_2)}\,\Big[ B_1 - B_2 + B_4 + B_5 + B_6 \Big]
    - {3\over 4 C_1}\,\Big[ B_1 - B_3 - 3 B_4 - B_6 \Big] \,,
    \nonumber\\
   && \\
   B &=& {3\over 4(C_1+C_2)}\,\Big[ B_1 - B_2 + B_4 + B_5 + B_6 \Big]
    + {1\over 4 C_1}\,\Big[ B_1 - B_3 - 3 B_4 - B_6 \Big] \,.
    \nonumber
\end{eqnarray}
Using the next-to-leading order expressions for the Wilson coefficients
given in (\ref{magic}), we find that
\begin{eqnarray}
   b(\mu) &=& {16\over\beta_0}\,\ln{\alpha_s(\mu)\over\alpha_s(m_Q)}
    + {\alpha_s(m_Q)\over\pi}
    - 8\,T_{34}\,{\alpha_s(m_Q)-\alpha_s(\mu)\over\pi} \,, \nonumber\\
   && \\
   B(\mu) &=& {16\over 9}\,C_{\rm mag}(\mu)\,\bigg\{ 1 + {3\over 2}\,
    T_{32}\,{\alpha_s(\mu)\over\pi} \bigg\} - {7\over 9}
    - \bigg( {41\over 27} + {8\over 3}\,T_{32} \bigg)\,
    {\alpha_s(m_Q)\over\pi} \,. \nonumber
\end{eqnarray}
{}From the fact that the physical decay constants $f_M$ in (\ref{fPfV})
must be $\mu$-independent, one can deduce the scale dependence of the
low-energy parameters. At leading order in $1/m_Q$, we recover the
well-known result that $C_1(\mu)\,F(\mu)$ is $\mu$-independent. At
order $1/m_Q$, the scale independent combinations are $G_1(\mu) -
\frac{1}{6}\bar\Lambda\,b(\mu)$ and $C_{\rm mag}(\mu)\,G_2(\mu) -
\frac{1}{12}\bar\Lambda\,B(\mu)$. Note that the structure of the
coefficients $b(\mu)$ and $B(\mu)$ is consistent with this requirement,
i.e., the $\mu$-dependent terms in $b(\mu)$ are independent of $m_Q$,
and those in $B(\mu)$ are proportional to $C_{\rm mag}(\mu)$. This
provides a nontrivial test of our results.\footnote{We have checked
that this ``consistency'' of our short-distance analysis with the
requirement of renormalizability of hadronic matrix elements is also
obeyed for $\bar B\to\pi\,\ell\,\bar\nu$ transitions, as well as for
the decay constants of excited mesons.}
As a consequence, we can define renormalization-group invariant, $\mu$-
and $m_Q$-independent low-energy parameters by
\begin{eqnarray}
   G_1^{\rm ren} &=& G_1(\mu) - {\bar\Lambda\over 6}\,\bigg\{
    {16\over\beta_0} \ln\alpha_s(\mu) + 8\,T_{34}\,
    {\alpha_s(\mu)\over\pi} \bigg\} \,, \nonumber\\
   &&\\
   G_2^{\rm ren} &=& \Big[ \alpha_s(\mu) \Big]^{-3/\beta_0}\,\bigg\{
    1 - S_{\rm mag}\,{\alpha_s(\mu)\over 4\pi} \bigg\}\,\bigg\{
    G_2(\mu) - {4\bar\Lambda\over 27}\,\bigg[ 1 + {3\over 2}\,T_{32}\,
    {\alpha_s(\mu)\over\pi} \bigg] \bigg\} \,. \nonumber
\end{eqnarray}
In terms of these parameters, one obtains, for instance,
\begin{equation}
   G_1(\mu) - {\bar\Lambda\over 6}\,b(\mu) = G_1^{\rm ren}
   + {\bar\Lambda\over 6}\,\bigg\{ {16\over\beta_0} \ln\alpha_s(m_Q)
   - \Big( 1 - 8\,T_{34} \Big)\,{\alpha_s(m_Q)\over\pi} \bigg\} \,,
\end{equation}
and a similar relation for the invariant combination involving
$G_2(\mu)$. These relations show explicitly the nonanalytic
$m_Q$-dependence of the $1/m_Q$ corrections to meson decay constants.

\bigskip
{\it Acknowledgments:\/}
Financial support from the BASF Aktiengesellschaft and from the German
National Scholarship Foundation is gratefully acknowledged. This work
was also supported by the Department of Energy, contract
DE-AC03-76SF00515.

\end{document}